\documentclass [12pt,a4paper] {article}
\usepackage{graphicx}
\usepackage{epsfig}

\usepackage[symbol]{footmisc}

\begin {document}

\hspace{10.5cm} November 2025
\bigskip

\begin{center}

{\bf{\Large Critical Phenomena in Hadronic and DIS Processes}}
\vspace{.5cm}

L.L.~Jenkovszky$^1$ and C.~Merino$^2$\\
\vspace{.5cm}

$^1$Bogolyubov Institute for Theoretical Physics (BITP)\\
National Academy of Sciences of Ukraine\\
14-b Metrolohichna str., Kiev 03143\\
Ukraine\\
%%%E-mail: jenk@bitp.kiev.ua

\vspace{.2cm}

$^2$Departamento de F\'\i sica de Part\'\i culas, Facultade de F\'\i sica \\
Instituto Galego de F\'\i sica de Altas Enerx\'\i as (IGFAE) \\
(Mar{\'i}a de Maeztu Unit of Excellence) \\
Universidade de Santiago de Compostela\\
Campus Universitario s/n, 15782 Compostela\\
Galiza, Spain \\
e-mail: carlos.merino@usc.gal

\vskip 0.5 cm

\end{center}

%\date{}

\begin{abstract}
We compare the critical phenomena (e.g. phase transitions, crossover) in proton-proton and proton-nucleus scattering
(larger systems), and in lepton-proton deep inelastic scattering (smaller systems).
%%&We compare the critical phenomena (e.g. phase transitions, crossover) in large (proton-proton, proton-nucleus scattering)
%%%and in small (lepton-proton deep inelastic scattering (DIS) systems).
\end{abstract}

\section{Collective effects in hadronic and DIS collisions}

Collective phenomena in proton-proton scattering (larger systems) and in deep-inelastic (DIS) lepton-proton scattering
(smaller systems)
have been extensively studied, and both similarities but also differences have been noticed between the two cases.

In principle, the transition from individual quark-quark, quark-hadron, hadron-hadron, hadron-nuclei, to lepton-quark, or
lepton-hadron collisions to collective phenomena is quite the same, but also significant differences can be identified
related to the fact that while in hadronic collisions two similar extended particles (hadrons or nuclei) interact, in DIS the
probe is a point-like particle (a lepton radiating a photon). Furthermore, one substantial difference regards the two formalisms
used to describe each case: hadronic collisions are studied in the c.m. (or lab.) system, but the DIS formalism is
intrinsically connected to the infinite momentum frame, $p_z\rightarrow\infty$.

However, the common crucial point is the creation in both cases of a new state of matter formed by a soup of quarks and gluons,
with saturation and transition from uncorrelated (leaving apart energy-momentum conservation) collisions of individual particles
to collective phenomena. We analyze the physics of these collective phenomenona in pp (pA, AA) and in ep (eA) scattering at high
energies and high virtualities of the probe, in the frame of the well known and stablished van der Waals (vdW) approach that may serve
as a clear and well defined bridge between the two scenarios. 

\section{Hadronic scattering}\label{Hadron}
   
The treatment of the van der Waals forces in hadronic systems has been consistently established in the litterature (see refs.~\cite{Fermi, LL, VAG1}).

The van der Waals equation of state is a model to describe the pressure function in equilibrium systems of particles with
both repulsive and attractive interactions that predicts a first-order liquid-gas phase transition and the corresponding critical
point~\cite{VAG1, VAG2, VAG3, VAG4, Vova}.

In the canonical ensemble (CE) the vdW equation of state has the following form:
\begin{equation}
p(T,n) = \frac{NT}{V-bN}-a\frac{N^2}{V^2}\equiv\frac{nT}{1-bn}-a n^2\;, 
\end{equation}
where $a>0$ and $b>0$ are the vdW parameters describing attractive and repulsive interactions, respectively, and $n\equiv N/V$ is
the particle number density. In order to apply the vdW equation of state to systems with variable number of particles one has to
consider the grand canonical ensemble (CGE), where the quantum statistics is easier to introduce, the vdW equation of state with Fermi
statistics being used to describe nuclear matter~\cite{Jenk1, Jenk2, Jenk3, Jenk4, Jenk5}.

The thermodynamics of nuclear matter and its application to heavy ion collisions has been studied already for more than forty years.
In particular, models employing a self-consistent mean field approach~\cite{MFA1, MFA2, MFA3, MFA4, MFA5, MFA6} are used to describe
the properties of nuclear matter.
The presence of the liquid-gas phase transition in nuclear matter has been experimentally detected~\cite{LGPT}, and direct measurements
of the nuclear caloric curve have been published~\cite{NCC1, NCC2}.

The vdW pressure isotherms in ($T$,$v$) and ($T$,$n$) coordinates ($v\equiv 1/n$) obtained in ref.~\cite{VAG1} are shown in Fig.~1 (extracted
from ref.~\cite{VAG1}). The
critical temperature is found to be $T_c\simeq19.7\ MeV$, close to the experimental estimates in references~\cite{ECT1, ECT2}.
At $T<T_0$ two phases appear: the gas and liquid phases separated by a first-order phase transition, the mixed phase region being shown
by horizontal lines in Fig.~1.a
and by shaded grey area in Fig.~1.b.
\begin{figure}[ht] 
\center{\includegraphics[width=.96\textwidth]{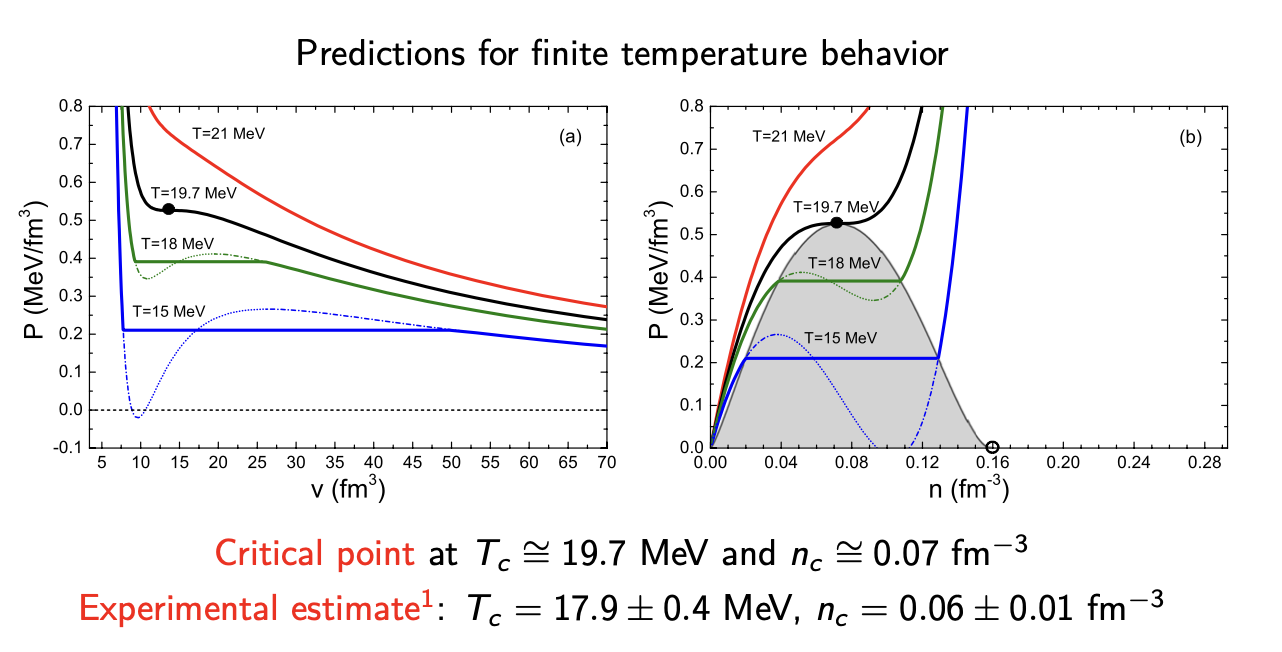}}
\caption{Pressure isotherm in (a) ($p$, $v$) and (b) ($p$, $n$), calculated in the quantum van der Waals equation of state 
(see details in ref.~\cite{VAG1}). The full circle on the $T=T_c$ isotherm corresponds to the critical point. Shaded grey area
in (b) accounts for the mixed phase region. This figure has been borrowed from ref.~\cite{VAG1}.
}
\label{Fig:alpha}
\end{figure}

Thus, the vdW equation with Fermi statistics applied to a system of interacting nucleons predicts a first-order liquid-gas phase
transition with a critical endpoint.
In highly inelastic hadronic collisions, once the hadron is broken by the interaction, the evolution through a stochastic process
of the wave function of a system of very many interacting particles would drive the inclusive spectrum to
the thermal distribution holding at the parton level~\cite{Krzywicki}.
Then, to deal with the problem of entropy production during the hadronization process, and based on observation
of Regge trajectories, possible entropy mechanisms far from equilibrium can be developed in terms of stochastic
dynamics~\cite{BiroSchramJenk2018}.

\section{Deep inelastic scattering}\label{Stat}

We will assume that in inclusive deep inelastic scattering (DIS),
or exclusive deeply virtual Compton scattering (DVCS),
the interior of a nucleon (or a nuclei) is seen as a thermodynamic
system that, as in the case of nuclear and heavy-ion collisions
bear collective (thermodynamic) properties governed by a relevant
equation of state (EoS). The idea that DIS structure functions (SF)
can be treated thermodynamically by means of statistical mechanics,
though not new, continues to attract attention, but several subtle
points remain still unclear, in particular in what it concerns to the
choice of the appropiate coordinate system and the corresponding
variables.

To set the Bjorken $x$ as the variable in the statistical distribution in DIS, instead of the
energy (or momentum), one has to introduce a proper dimensional parameter related to the change of the
coordinate system
\begin{eqnarray}
\exp\left(\frac{-x}{\bar{x}}\right)\;,\ with\ \bar{x}=x\cdot\left(\frac{1-k_T^2}{x^2m^2}\right)\;,  
\end{eqnarray}

\noindent
where $k_T$ is the quark transverse momentum and $m$ is the mass of the proton.

On the other hand, also here, as in any statistical distribution, a temperature $T$ exists, though while
a limited temperature is typical of the hadronic phase~\cite{Hagedorn}, now that $T$ will be the temperature
of the partonic gas (liquid/fluid phase), which can be heated indefinitely, and consequently, it can not be
limited.

The experimentally observed saturation of the DIS structure functions indicates a phase transition from the partonic
gas (high and intermediate $x$) to a fluid (small $x$). Thus the partonic matter in a nucleon (or nucleus) transforms
from a almost perfect gas (Bjorken scaling region) to a fluid (the logarithmic scaling violation is replaced by a power law).
To identify the presence of these two regions: the Bjorken scaling (dilute) region, and the (dense) region beyond it, with
a intermediate mixed region, the relevant variables are the parton fraction of the nucleon momentum, $x$, and the virtuality
of the incident photon, $Q^2$. As in the case of hadronic and nuclear collisions, also in DIS (i.e. for smaller systems),
the presence of the two (gaseous and fluid) phases can be described by a van der Waals equation of state. 

\subsection{Statistical models of DIS structure functions}

Deep inelastic scattering structure functions are sums of a singlet (S) and a non-singlet (NS) terms,
\begin{eqnarray}
F_2(x,Q^2)=F_2^S(x,Q^2)+F_2^{NS}(x,Q^2)\;, 
\end{eqnarray}

\noindent
each the product of a low-$x$ ($\sim x^{\alpha}$) and a high-$x$ $(1-x)^n$ factors:
\begin{eqnarray}
F_2^S(x,Q^2)=A_0\left(\frac{Q^2}{Q^2+a}\right)^{1+\Delta(Q^2)}x^{-\Delta(Q^2)}(1-x)^{n(Q^2)+4}\;,\
\Delta(Q^2)=\Delta_0\left(1+\frac{bQ^2}{Q^2+c}\right)\;. 
\end{eqnarray}
This model is applicable in the Regge domain of small and intermediate values of $Q^2$ (see details
of the model in ref.~\cite{BertiniJenk}. At hihger vistualities, Regge behaviour should be replaced by the effects
of QCD evolution, as indicates the drastic increase of $F_2(x,Q^2)$ in $1/x$ with increasing $Q^2$
(new dynamical regime). Although in DIS (for virtual particles) there is no analogue to the Froissart
bound, also of change of dynamical scenario should occur since with the power-like scaling violating rise of
$F_2(x,Q^2)$ the number of partons increases and their volume tends to eceed taht of the nucleon
(gluon saturation), leading to partons recombination or condensation (in terms of statistical physics).

In ref.~\cite{DesgrolardJenk} a model of the structure functions based on the interpolation between the Regge and DGLAP regimes
and realizing the onset of saturation was presented. and used~\cite{JenkNagy} to match the saturation region in DIS
with the predicted phase transition.

Since the saturation phenomena occur in the small-$x$ region, which is gluon-dominated,
one is mainly interested in that component of the SF:
\begin{eqnarray}
F_2\sim x\cdot G(x,Q^2)\;,
\end{eqnarray}

\noindent
where $G$ is the gluon (singlet) distribution function.

Thus we will, also for simplicity, focus on the small-$x$ singlet (gluon) component
of the SF, the extension to low-$x$ and/or the non-singlet (valence quark)
contributions being straightforward
\begin{equation}
x\cdot G(x,Q^2)\sim {\frac{X_0x^b}{\exp{[(x-X_0)/\bar x]}+1}}\;,
\label{xG}
\end{equation}
where $x$ is the Bjorken (light-cone) variable, $X_0$ is the chemical
potential, that for the gluon component can be set zero, and $\bar x$
is interpreted as the temperature inside the proton.

In the litterature~\cite{Bhalerao, Mac, Cleymans1} the dimensional energy $E$ (or momentum $k$)
variable has been used in the statistical model of the SF, instead of $x$
as in Eq.~(\ref{xG}). This is not a trivial kinematical problem,
since thermodynamics implies the presence of the dimensional
temperature in the statistical distribution like $k/T$ (be it of
the Fermi-Dirac, Bose-Einstein, or Boltzmann type), while the
appearance of $x$ as in Eq.~(\ref{xG}) needs some extra modification.
This can be circumvent (see ref.~\cite{Cleymans2}) by using
a dimensionless temperature $\bar x=2T/m$, where $m$ is the
proton mass, which is a consequence of the transition from the
rest frame to the infinite-momentum frame (IMF). Accordingly,
\begin{equation}
G(x)\sim\exp\Bigl{(-\frac{mx}{2T}}\Bigr)\;.
\label{G}
\end{equation}
The Boltzmann factor in the denominator of Eq.~(\ref{xG}) can
mimic the large-$x$ $(1-x)^n$ factor in the SF, although it
should be reconciled also with the quark counting rules appearing
to the power $n$.

One last point is also connected with the EoS expected from the
statistical distribution of the type in Eq.~(\ref{xG}). Let us remember
that for an ideal gas of particles
\begin{equation}
P(T)=\int_0^\infty k d^3k \exp(-k/T)\;, \nonumber
\end{equation}
that due to radial symmetry can be rewritten as
\begin{equation}
\int_0^\infty k^3 dk \exp(-k/T)\;, \nonumber
\end{equation}
and by making the change of variable $y=k/T$ one trivially arrives at
the Stafan-Boltzmann (S-B) EoS:
\begin{equation}
P\sim T^4\;. \nonumber
\end{equation}
This fact has a physical interpretation: the large-$x$ component of the SF
corresponds to a dilute perfect gas of partons. The low-$x$ factor
in Eq.~(\ref{xG}) will affect the ideal S-B EoS only
when it will be written as
\begin{equation}
(x/\bar x)^b=({mx\over{2T}})^b\;, \nonumber
\end{equation}
instead of $x^b.$ As a consequence, the
ideal S-B EoS will be modified to
\begin{equation}
P(T)\sim T^{4+b}\; .
\label{P}
\end{equation}
The relative contribution of this correction is negligible at small $x$,
but it increases with $Q^2$ and decreasing $x$, resulting in a gas-liquid
phase transition.

Usually the $Q^2$ dependence is neglected,
either for simplicity, or conceptually, by assuming that the
statistical approach applies to the SF for some fixed input
value of $Q^2$ from which it evolves according to the DGLAP
equation. We do not exclude high $Q^2$ evolution of the SF,
however with the following caveats:

a) the structure functions show strong $Q^2$ dependence, already
at low $x$, below the perturbative DGLAP domain; b) at large
$Q^2$, instead of the monotonic DGLAP evolution, and due to the
proliferation of partons, the inverse process of their
recombination is manifest, this process being essential in our
interpretation of the saturation as a gas-liquid phase transition
(see next section). Thus we prefer to keep explicit $Q^2$
dependence for all $x$ and $Q^2$. This dependence is mild in
the gaseous region of point-like partons (at large $x$), but it becomes
significant towards the saturation region (depending on both $x$ and $Q^2$),
where the point-like partons are replaced by finite-size droplets
of the partonic fluid. In the next section we will treat this transition
by using the classical van de Waals equation.

\subsection{Gas-Fluid phase transition in the van der Waals equation of
state}\label{IV}

Once the statistical properties of the SF have been defined, we now proceed
to write an equation of state (EoS) describing the transition from a parton
gas to the partonic liquid, via a mixed foggy phase. To do this we use the
van der Waals equation (see~\cite{Fermi, LL}):
\begin{equation}
(P+N^2a/V^2)(V-Nb)=NT\;,
\label{VdW}
\end{equation}
where $a$ and $b$ are parameters depending on the properties
of the system, $N$ is the number of particles, and $V$ is the
volume of the container:
\begin{equation}
V(s)=\pi R^3(s)\;,\ \nonumber
\end{equation}

\noindent
and
\begin{equation}
R(s)\sim\ln s \nonumber
\end{equation}
is the nucleon radius for our case. For point-like particles
(perfect gas), $a=b=0,$ and Eq.~(\ref{VdW}) reduces to
\begin{equation}
pV=NT\; , \nonumber
\end{equation}
and, since $N/V \sim T^3$, we get in this approximation
\begin{equation}
p\sim T^4\; , \nonumber
\end{equation}
to be compared with $p\sim T^{(4+b)}$ in Eq.~(\ref{P}).

On the other way, Eq.~(\ref{VdW}) can be also written as~\cite{Fermi}
\begin{equation}
(P+a/V^2)(V-b)=RT\; , \nonumber
\end{equation}
or, equivalently
\begin{equation}
P=\frac{RT}{V-b}-\frac{a}{V^2}\; . \nonumber
\end{equation}
The parameter $b$ is responsible for the finite dimensions of
the constituents, related to $1/Q$ in our case, and the term
$a/V^2$ is connected to the (long-range) forces between the
constituents. From this cubic equation in $V$ one finds~\cite{Fermi}
that the critical values for the main magnitudes
$V=V_c$, $P=P_c$, and $T=T_c$,
can be written in terms of the parameters $a$ and $b$ as:
\begin{equation}
V_c=3b,\ \ p_c=a/(27b^2)\ \  T_c=8a/27Rb\; . \nonumber
\end{equation}
The number of particles $N(s)$ can be calculated~\cite{Jenk1, JStr} as
\begin{equation}
N(s)=\int_0^1dxF_2(x,Q^2)\; , \nonumber
\end{equation}
where $F_2(x,Q^2)$ is the nucleon structure function, measured in DIS.

Now, we remind the kinematics:
\begin{equation}
s=Q^2(1-x)/x+m^2\; , \nonumber
\end{equation}
which at small $x$ reduces to $s\approx Q^2/x$. The radius of the constituent as seen
in DIS is $r_0\sim 1/Q$, hence its two-dimensional volume is $\sim Q^{-2}$.

By introducing the so-called reduced volume, pressure and temperature:
\begin{equation}
{\cal P}=P/P_c,\ \ {\cal V}=V/V_c=\rho_c/\rho,\ \  {\cal T}=T/T_c\; , \nonumber
\end{equation}
the van der Waals equation~(\ref{VdW}) can be rewritten as
\begin{equation}
\Bigl({\cal P}+3/{\cal V}^3\Bigr)\Bigl({\cal V}-1/3\Bigr)=8{\cal T}/3 \; .
\label{VdWCr}
\end{equation}

Note that Eq.~(\ref{VdWCr}) contains only numerical constants, and
therefore it is universal.
States of various substances with the same values of
${\cal P},\  {\cal V}$ and ${\cal T}$ are called
corresponding states and equation~(\ref{VdWCr}) is known as the
van der Waals equation for corresponding states. The universality
of the liquid-gas phase transition and the corresponding principle are
typical for any system with short-range repulsive and long-range
attractive forces. This property is shared both by ordinary liquids
and by nuclear matter~\cite{Jaqaman1}.

Let us show now two examples of EoS, one based on the Skyrme effective
interaction and finite-temperature Hartree-Fock theory, and the second
one being the van der Waals EoS.

For one EoS based on the Skyrme effective interaction and finite-temperature
Hartree-Fock theory, in ref.~\cite{Jaqaman1, Jaqaman2} one uses the EoS
\begin{equation}
P=\rho kT-a_0\rho^2+a_3(1+\sigma)\rho^{(2+\sigma)}\; ,
\label{Jaq1}
\end{equation}
where $\rho=N/V$ is the density and $a_0,\ \ a_3$ and $\sigma$ are
parameters, $\sigma=1$ accounting for the usual Skyrme interaction. 

According to the law of the corresponding states Eq.~(\ref{Jaq1})
is universal for scaled (reduced) variables, for which, with $\sigma=1$, it becomes
\begin{equation}
P=3{\cal  T}/{\cal V}-3/{\cal V}^2+1/{\cal V}^3\; . \nonumber
\end{equation}

This EoS is to be compared with the van der Waals EoS of the second example
\begin{equation}
P=8{\cal  T}/(3{\cal V}-1)-3/{\cal V}^2\; . \nonumber
\end{equation}

If we now write the van der Waals EoS in the form
\begin{eqnarray}
P(T;N,V)&=&-\Bigl(\frac{\partial F}{\partial
V}\Bigr)_{TN} \nonumber \\
&=&\frac{NT}{V-bN}-a\Bigl(\frac{N}{V}\Bigr)^2=
\frac{nT}{1-bn}-an^2\; ,
\label{VdW11}
\end{eqnarray}
with $n=N/V$ the particle number density, $a$ the strength
of the mean-field attraction, and $b$ governing the short-range
repulsion. We identify the particle number density with the SF
$F_2(x,Q^2)$.

\begin{figure}[h]
\begin{center}
\hspace{-1.cm}
\includegraphics[width=0.8\textwidth,angle=0]{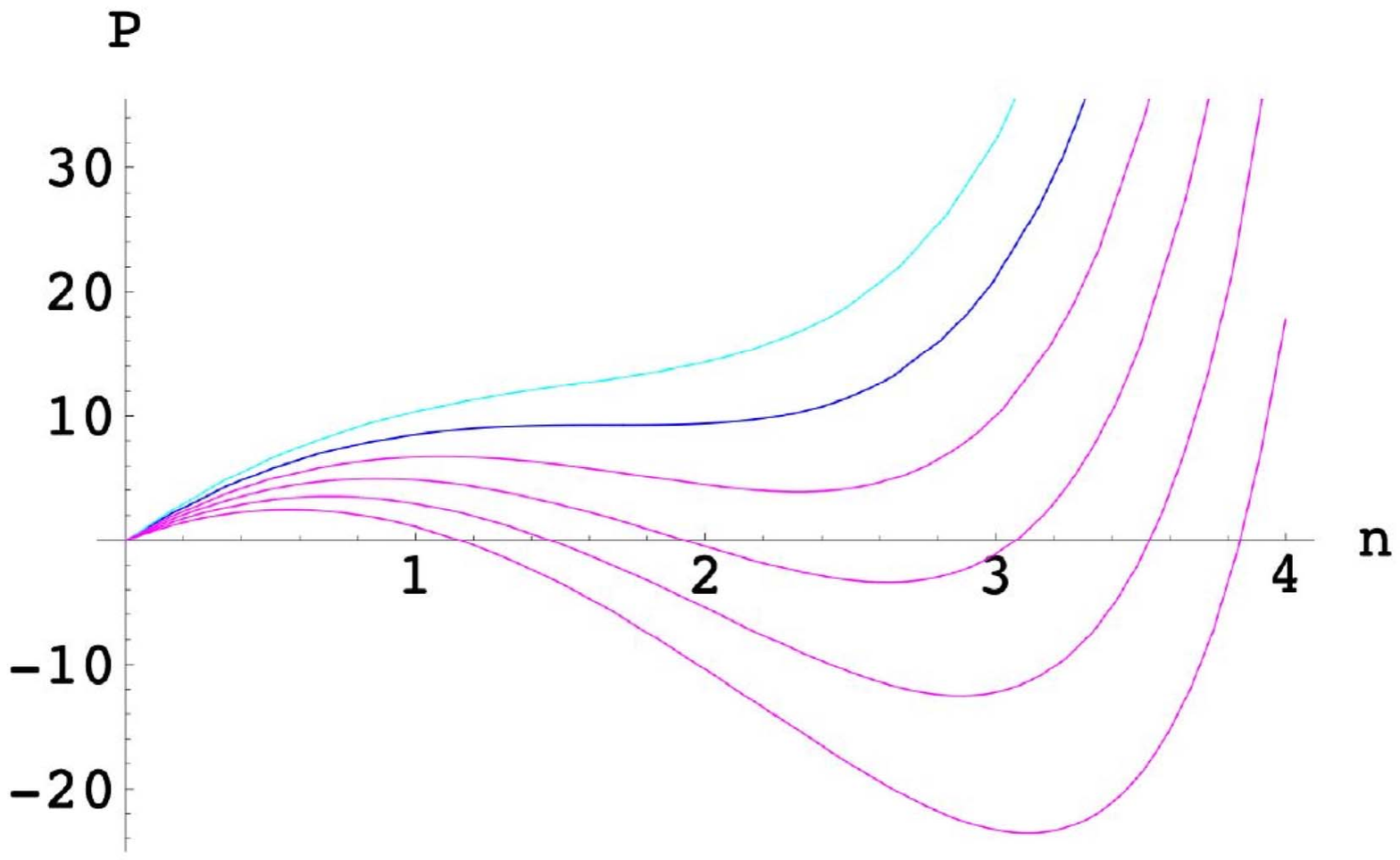}
\caption{\small \it {The pressure-to-density dependence
calculated, in arbitrary units, from Eq.(\ref{VdW11}).}}
\label{Fig:VdW1}
\end{center}
\end{figure}

Fig.~\ref{Fig:VdW1} shows the
pressure-density dependence calculated from Eq.~(\ref{VdW11}),
with $a=5\ GeV^{-2}$ and $b=0.2\ GeV^{-3}$.

Notice that while Eq.~(\ref{VdW11}) is sensitive to $b$ (short-range
repulsion), it is less so to $a$ (long-range attraction). Representative
isotherms are shown in this figure, the dark blue line (second from the top)
being the critical one, $T_c=8a/27b$. Above this temperature (top
line, in pale blue), the pressure rises uniformly with density,
corresponding to a single thermodynamical state for each $P$ and
$T$. For subcritical temperatures, $0<T<T_c$, (red
lines), on the contrary, the function $P(n)$ presents a maximum followed by a minimum.
Below the critical value $P_c$, three density regimes exist. The smallest density region
lies in the gaseous phase below the spinodal region, while the highest densities lie in the
liquid phase, above the spinodal region.
The coexistence phase can be determined by a Maxwell construction.

Experimentally, the onset of the gas-fluid phase transition may be verified by the observed spectrum
of the produced particles. The $p_{\perp}$ distribution of particles produced from the dilute
gaseous state (below the saturation region), can be computed perturbatively, while beyond the
saturation border line the produced particles result from the collisions of a very large number of
constituents, as in the color glass condensate~\cite{CGC1,CGC2}. Whatever the details of
the phase transition, the significant point is the existence of a dense partonic medium. Experimental
data on DIS and on high-energy proton-nucleus and heavy-ion collisions show that the partonic matter
may appear as a fluid, the color glass condensate. The nature of strongly interacting matter under
extreme conditions should be universal, be it produced in hadron-hadron, hadron-nucleus, nucleus-nucleus,
heavy-ion collisions, or in deep inelastic lepton-hadron scattering.

To map the saturation region in DIS onto the spinodal region in the VdW EoS, that is,
to make the correspondence between the EoS with its variables $P,\ T,\
\mu$, {\it etc.} to the experimental observables, depending on the reaction
kinematics, is the most delicate and complicated point in the
thermodynamical description of any high-energy collisions, in particular of DIS.

It requires further cautious studies and accurate numerical
tests. Attempts to link two different approaches,
one based on the $S$ matrix (scattering amplitude, cross
sections) and the other one on their collective properties
(statistical mechanics, thermodynamics, equation of state), to hadron dynamics,
have already been proposed (see~\cite{Jenk1} and references therein).

\section{Summary}

The description, formalism and most characteristic features of critical phenomena (phase transitions)
in hadronic processes and deep inelastic scattering are presented, and the formalism and most characteristic features
in both cases are exposed and compared by stressing the main similarities and differences.

In hadronic collisions, the vdW equation with Fermi statistics applied to a system of interacting nucleons predicts
a first-order liquid-gas phase transition with a critical endpoint. In highly inelastic hadronic collisions, once the
hadron is broken by the interaction, the inclusive spectrum would reflect the thermal distribution holding at the parton
level. Then, to deal with the problem of entropy production during the hadronization process, and based on observation
of Regge trajectories, possible entropy mechanisms far from equilibrium can be developed in terms of stochastic
dynamics.

In DIS, the experimentally observed saturation of the structure functions indicates a phase transition from the partonic
gas (high and intermediate $x$) to a fluid (small $x$). Thus the partonic matter in a nucleon (or nucleus) transforms
from a almost perfect gas (Bjorken scaling region) to a fluid (the logarithmic scaling violation is replaced by a power law).
As in the case of hadronic and nuclear collisions, also in DIS (i.e. for smaller systems),
the presence of the two (gaseous and fluid) phases can be described by a van der Waals equation of state.
To get the mapping of the
saturation region in DIS onto the spinodal region in the van der Waals equation of state further detailed studies and
advanced numerical calculations are needed.

Experimentally, the onset of the gas-fluid phase transition may be verified by the observed spectrum
of the produced particles. The $p_{\perp}$ distribution of particles produced from the dilute
gaseous state (below the saturation region), can be computed perturbatively, while beyond the
saturation border line the produced particles result from the collisions of a very large number of
constituents, as in the color glass condensate.

Whatever the details of the phase transition, the significant point is the existence of a dense partonic medium.
Experimental data on DIS and on high-energy proton-nucleus and heavy-ion collisions show that the partonic matter
may appear as a fluid, the color glass condensate. The nature of strongly interacting matter under
extreme conditions should be universal, be it produced in hadron-hadron, hadron-nucleus, nucleus-nucleus,
heavy-ion collisions, or in deep inelastic lepton-hadron scattering.

\end{document}